\documentclass[10pt,twocolumn,letterpaper]{article}

\usepackage{cvm}
\usepackage{times}
\usepackage{epsfig}
\usepackage{graphicx}
\usepackage{amsmath}
\usepackage{amssymb}

\usepackage{booktabs}
\usepackage{multirow}
\usepackage{array,makecell,graphicx}

\usepackage{array}
\usepackage{makecell}

\newcolumntype{C}[1]{>{\centering\arraybackslash}p{#1}}

\usepackage[pagebackref=true,breaklinks=true,letterpaper=true,colorlinks,bookmarks=false]{hyperref}

\newcommand{\keywords}[1]{{\bf \emph{Keywords: #1}}}

\cvmfinalcopy

\def\cvmPaperID{294} 

\ifcvmfinal\fi
\begin{document}

\title{ORACLE: Orchestrate NPC Daily Activities using Contrastive Learning with Transformer-CVAE}

\author{Seong-Eun Hong\\
Korea University\\
Seoul, South Korea\\
{\tt\small seong\_eun@korea.ac.kr}
\and
JuYeong Hwang\\
Korea University\\
Seoul, South Korea\\
{\tt\small 05judy02@korea.ac.kr}
\and
RyunHa Lee\\
Korea University\\
Seoul, South Korea\\
{\tt\small bluerish@korea.ac.kr}
\and
HyeongYeop Kang\thanks{Corresponding author}\\
Korea University\\
Seoul, South Korea\\
{\tt\small siamiz\_hkang@korea.ac.kr}
}

\maketitle

\begin{abstract}
    The integration of Non-player characters (NPCs) within digital environments has been increasingly recognized for its potential to augment user immersion and cognitive engagement. 
  The sophisticated orchestration of their daily activities, reflecting the nuances of human daily routines, contributes significantly to the realism of digital environments. 
  Nevertheless, conventional approaches often produce monotonous repetition, falling short of capturing the intricacies of real human activity plans. 
  In response to this, we introduce ORACLE, a novel generative model for the synthesis of realistic indoor daily activity plans, ensuring NPCs' authentic presence in digital habitats. 
  Exploiting the CASAS smart home dataset's 24-hour indoor activity sequences, ORACLE addresses challenges in the dataset, including its imbalanced sequential data, the scarcity of training samples, and the absence of pre-trained models encapsulating human daily activity patterns. 
  ORACLE's training leverages the sequential data processing prowess of Transformers, the generative controllability of Conditional Variational Autoencoders (CVAE), and the discriminative refinement of contrastive learning. Our experimental results validate the superiority of generating NPC activity plans and the efficacy of our design strategies over existing methods.
\end{abstract}


\section{Introduction}

Non-player characters (NPCs) are fundamental components of digital environments such as video games and virtual reality (VR), acting as interactive agents that enhance user experiences. Beyond simple task execution, NPCs that exhibit realistic daily routines contribute to deeper immersion by making digital worlds feel more lifelike and dynamic~\cite{hasani2021immersive, mateas2001oz}. However, generating diverse, natural, and contextually coherent daily activity sequences for NPCs remains a significant challenge~\cite{avradinis2013behavior}. 

Traditional approaches to NPC activity modeling have relied on scripted rule-based systems, hierarchical task networks (HTN)~\cite{bogdanovych2017plan}, and belief-desire-intention (BDI) models~\cite{de2005motivational}. While these methods allow for structured behavior generation, they suffer from limited variability and require extensive manual design to adapt to new environments. More recently, learning-based sequence modeling techniques have been explored, leveraging recurrent and generative models to infer NPC activity sequences from data. Despite these advancements, existing approaches still struggle with three challenges. First, it is challenging to capture long-term dependencies in daily routines, as many models rely on recurrent architectures such as LSTMs, which suffer from vanishing gradients and limited memory when processing long sequences, leading to short-term planning that fails to account for realistic, temporally coherent activity patterns. Second, generating diverse yet realistic schedules remains a challenge, since deterministic models often produce repetitive outputs and fail to introduce controlled variability. Many sequence modeling approaches optimize for likelihood maximization, which biases them toward generating frequent patterns rather than exploring underrepresented but plausible variations in human activity. Third, ensuring coherence when integrating partially defined activity constraints is difficult, as traditional autoregressive and transformer-based models struggle to reconcile predefined activities with generated sequences. These models lack mechanisms to balance adherence to constraints while preserving natural transitions between activities, leading to inconsistencies such as unrealistic gaps or abrupt shifts in daily routines~\cite{koushik2023activity}.

In parallel, recent work has explored using large language models (LLMs) to plan multi-step behaviors for simulated or embodied agents. However, LLM-based pipelines typically impose substantial inference latency and GPU cost, and reliance on commercial APIs can further introduce per-request fees and governance constraints, which limit adoption in resource-constrained or latency-sensitive industrial settings. Moreover, while LLMs encode broad world knowledge, they are not trained on activity-plan–specific data (e.g., 24-hour routine corpora), often yielding generic or inconsistent schedules unless heavily fine-tuned. These considerations motivate the need for a compact, domain-tailored generative model for activity planning.

To address these limitations, we introduce \textbf{ORACLE} (\textbf{OR}chestrate NPC Daily \textbf{A}ctivities using \textbf{C}ontrastive \textbf{L}earning with CVA\textbf{E}), a generative framework that learns to synthesize realistic human-like daily routines for NPCs. ORACLE is trained on the \textbf{CASAS smart home dataset}~\cite{cook2019human,cook2012casas}, which provides 24-hour indoor activity sequences. Unlike previous sequence modeling approaches, which primarily focus on direct sequence prediction, ORACLE incorporates. 1) A latent variable model to introduce diversity in generated schedules while maintaining coherence. 2) A contrastive learning framework to refine the generative process, ensuring activity sequences adhere to realistic patterns. 3) A conditional generative structure that supports both full-schedule generation and augmentation of partially defined schedules.  

Unlike deterministic sequence generation methods, ORACLE \textbf{leverages probabilistic modeling} to create more flexible and adaptable schedules, reducing the tendency for repetitive or unrealistic behavior. The contrastive learning mechanism further enhances the model’s ability to recognize and correct \textbf{implausible activity sequences}, ensuring that generated NPC routines remain \textbf{coherent and lifelike}.  

In summary, the contributions of this paper are as follows:
\begin{itemize}
\item We introduce a generative model for NPC daily activity synthesis, integrating latent variable modeling and contrastive learning to enhance sequence diversity and realism.  
\item Our model is flexible, supporting both full-schedule generation and augmentation of partially defined schedules, making it adaptable to different application scenarios.  
\item We propose a contrastive learning framework tailored to human activity modeling, improving the model’s ability to generate contextually coherent daily routines.  
\item We present extensive experiments, including comparisons with existing sequence modeling approaches, a user study, and an ablation analysis, to validate the effectiveness of our approach.  
\end{itemize}

\section{Related work}
\subsection{Plausible NPC Activity Planning}
The development of NPCs that inhabit the digital realm like humans spans the fields of AI, gaming, and VR. Achieving this involves either assigning predefined activities to NPCs or enabling them to autonomously decide their actions based on their goals and environmental context.

Plan-based methodologies allow NPC to determine actions in alignment with their objectives and the current environmental conditions. 
STRIPS~\cite{fikes1971strips}, an early AI planning system, introduces a formal language for planning, defining actions, and state transitions. 
STRIPS is enhanced by Planning Domain Definition Language (PDDL)~\cite{aeronautiques1998pddl,fox2003pddl2} to cover more complex scenarios.
In the realm of video gaming, Goal-Oriented Action Planning (GOAP)~\cite{jeff2003applying,anderson2010developing,robertson2014review} enable NPCS to devise dynamic strategies tailored to their immediate circumstances, while Hierarchical Task Network (HTN)~\cite{ghallab2004automated, kelly2008offline} decomposes tasks into subtasks to create actionable plans.
However, these methods often lead to predictable and overly deterministic NPC behavior, reducing their realism.

Motivation-based methodologies use internal drives, such as hunger, safety, and social needs, to generate more varied and sometimes unpredictable NPC behaviors. These approaches emulate human biological and psychological imperatives to create realistic activity plans~\cite{chen2001equipping, de2005motivational, krumpelmann2011motivating, avradinis2013behavior, gramoli2021needs}.
The Beliefs-Desires-Intentions (BDI) model~\cite{bratman1987intention, dignum2000towards, parsons2000approach, de2020bdi, gramoli2022control} further integrates human-like cognitive processes into NPC design, guiding their actions through a matrix of beliefs and intentions. 
However, building and maintaining a BDI system is labor-intensive and computationally demanding, particularly in balancing diverse and conflicting intentions.

In recent advancements, learning-based methodologies have emerged. 
Contemporary research increasingly adopts reinforcement and deep learning techniques, empowering virtual agents to refine their decision-making through accumulated experiences~\cite{mccall2020artificial,jang2021deep,johansen2022towards,maroto2023biologically,park2023generative}.
Despite the potent capabilities of these methods, they require significant computational resources and extensive datasets for training. 
Particularly, the scarcity of comprehensive datasets recording human activity in naturalistic settings poses a substantial bottleneck~\cite{vanhulsel2009simulation,cook2019human,koushik2023activity}.

\subsection{Sequential data analysis}
The primary focus of sequential data analysis is developing models adept at delineating the temporal dynamics and dependencies inherent in data sequences. 
Historically, techniques such as the Hidden Markov Model (HMM)~\cite{baum1966hmm} and dynamic time warping~\cite{Bellman1959DTW} have been extensively studied.

Sequential data analysis has been substantially evolved with the advent of Natural Language Processing (NLP) and deep learning methodologies.  
Recurrent Neural Networks (RNN)~\cite{rumelhart1986backpropagation}, including variants such as Long Short-Term Memory (LSTM)~\cite{Hochreiter1997LSTM} networks and Gated Recurrent Unit (GRU)~\cite{cho2014gru, chung2014gru}, emerged as foundational models for processing sequential data. 
These architectures demonstrated exceptional capability in capturing long-range dependencies, thereby significantly advancing tasks such as machine translation~\cite{cho2014gru}, text generation~\cite{ranzato2015sequence}, and speech recognition~\cite{graves2013speech}. 

Despite their success, RNN and LSTM encountered notable training challenges over extended sequences, attributed primarily to vanishing and exploding gradient issues~\cite{hochreiter1998vanishing}. RNN and LSTM have been widely used to deal with sequence data. Koushik~\cite{koushik2023activity} conducted an activity-based travel demand modeling using bidirectional LSTM~\cite{graves2005framewise}.
The introduction of the Transformer~\cite{NIPS2017_transformer} model innovatively circumvented these challenges by leveraging self-attention mechanisms, enabling the parallel processing of sequence data and markedly enhancing model efficiency. 
Consequently, the Transformer architecture has become a cornerstone in the field, with foundational models such as  Bidirectional Encoder Representations from Transformers (BERT)~\cite{devlin2018bert}, Generative Pre-trained Transformer (GPT) ~\cite{radford2018gpt}, and their variants establishing new standards across a wide range of sequential data analysis applications. 

In parallel, there has been an increasing interest in leveraging generative models for sequential data analysis. 
Variational Autoencoder (VAE)~\cite{kingma2022VAE}, known for their proficiency in learning latent data representations, have shown promise in generating high-quality, diverse sequences that closely mimic the statistical properties of real-world data. 
Bowman~\cite{bowman2016generating} combine LSTM and VAE to provide diversity in sequence data as a generative model.
Additionally, Conditional variational Autoencoder (CVAE) achieves controllable generation through exploiting learned latent representations~\cite{zhao-etal-2017-learning}. Le Fang~\cite{fang2021transformer} suggests Transformer-based CVAE to bring the advantages of Transformer and CVAE.

On the other hand, contrastive learning has emerged as a pivotal technique in large-scale unsupervised learning, focusing on learning feature spaces by bringing similar samples (positive samples) closer together and pushing dissimilar ones (negative samples) apart~\cite{he2020momentum,chen2020simple}.
This method has been particularly successful in representation learning for sequential data~\cite{liu2021simcls}.

\begin{table}[tb]
  \centering
  \hspace*{-0.25cm}
  \setlength{\tabcolsep}{7pt}
  {\footnotesize
  \begin{tabular}{@{\hspace{3.2pt}}p{0.175\linewidth}|@{\hspace{3.2pt}}p{0.7509\linewidth}}
    \toprule
    Categorized Classes & \multirow{2}{*}{Criteria} \\
    \midrule
    Sleep & 5 to 12 hours a day, Once a day\\[0.2ex]
    Outing & Up to 12 hours a day, No restrictions \\[0.2ex]
    Rest & Up to 12 hours a day, No restrictions\\[0.2ex]
    Work & Up to 12 hours a day, No restrictions\\[0.2ex]
    Hygiene & Up to 1.5 hours per occurrence, At least once a day \\[0.2ex]
    Toilet & Up to 0.5 hours per occurrence, At least once a day \\[0.2ex]
    Dress & Up to 1 hour per occurrence, At least once a day \\[0.2ex]
    Cook & Up to 2 hours per occurrence, No restrictions\\[0.2ex]
    Meal & Up to 2 hours per occurrence, At least once a day\\[0.2ex]
    Chore & Up to 2 hours per occurrence, At least once a day\\[0.2ex]
    Snack & Up to 2 hours per occurrence, No restrictions\\[0.2ex]
    Medicine & Up to 0.3 hours per occurrence, No restrictions\\
  \bottomrule
  \end{tabular}
  \caption{The criteria for data filtering.}
  \label{tab:Criteria}
  }
\end{table}

\section{Dataset}

To train ORACLE on human activity planning within indoor home environments, we use the CASAS smart home dataset, which consists of two subsets: the Apartment dataset and the Home dataset. These datasets contain continuous 24-hour activity sequences recorded through ambient sensors installed in residential settings. The Apartment dataset includes 42 activity classes, while the Home dataset contains 46 classes. Data was collected from 20 participants (8 males, 12 females, aged 21-62, $\mu$ = 33), resulting in a total of approximately 57,216 hours of recorded activity in the Apartment dataset and 26,112 hours in the Home dataset.

Despite the extensive data collection, several challenges arise:  
1) Unlabeled periods categorized as \textit{Other activity}, reducing the reliability of activity segmentation.  
2) Severe class imbalance, where activities such as \textit{Entertain Guests} appear over 1.27 million times, while \textit{Exercise} appears only 93 times.  
3) Incomplete 24-hour cycles due to sensor interruptions, leading to inconsistencies in activity sequences.

\begin{table}[t!]
  \centering
  \begin{tabular}{p{0.37\linewidth}|p{0.50\linewidth}}
    \toprule
    Major activity classes & Original activity classes\\
    \midrule
    \addlinespace
    Sleep & sleep, sleep out of bed, go to sleep, nap, wake up \\
    Outing & leave home, enter home, step out, single leave, single enter, staff leave, staff enter, single step out \\
    Rest & watch TV, entertain guests, read, relax, phone, exercise, pet activity, sew \\
    Work & work, work on computer, work at table, work at desk \\
    Hygiene & personal hygiene, groom, bathe, shower \\
    Toilet & toilet, bed toilet transition \\
    Dress & dress \\
    Cook & cook, cook breakfast, cook lunch, cook dinner, \\
    Meal & eat breakfast, eat lunch, eat dinner \\
    Chore & wash dishes, wash breakfast dishes, wash lunch dishes, wash dinner dishes, laundry, housekeeping, put groceries away \\
    Snack & drink, eat \\
    Medicine & take medicine, morning meds, evening meds \\
  \bottomrule
  \hline
  \end{tabular}
  \caption{42 activity classes in Apartment dataset and 46 activity classes in Home dataset are merged into 12 major activities.}
  \label{tab:majorclassification}
\end{table}

\begin{figure*}[hbtp]
    \setlength{\abovecaptionskip}{1pt}
    \setlength{\belowcaptionskip}{-80pt}
    \centering 
    \includegraphics[width=\textwidth]{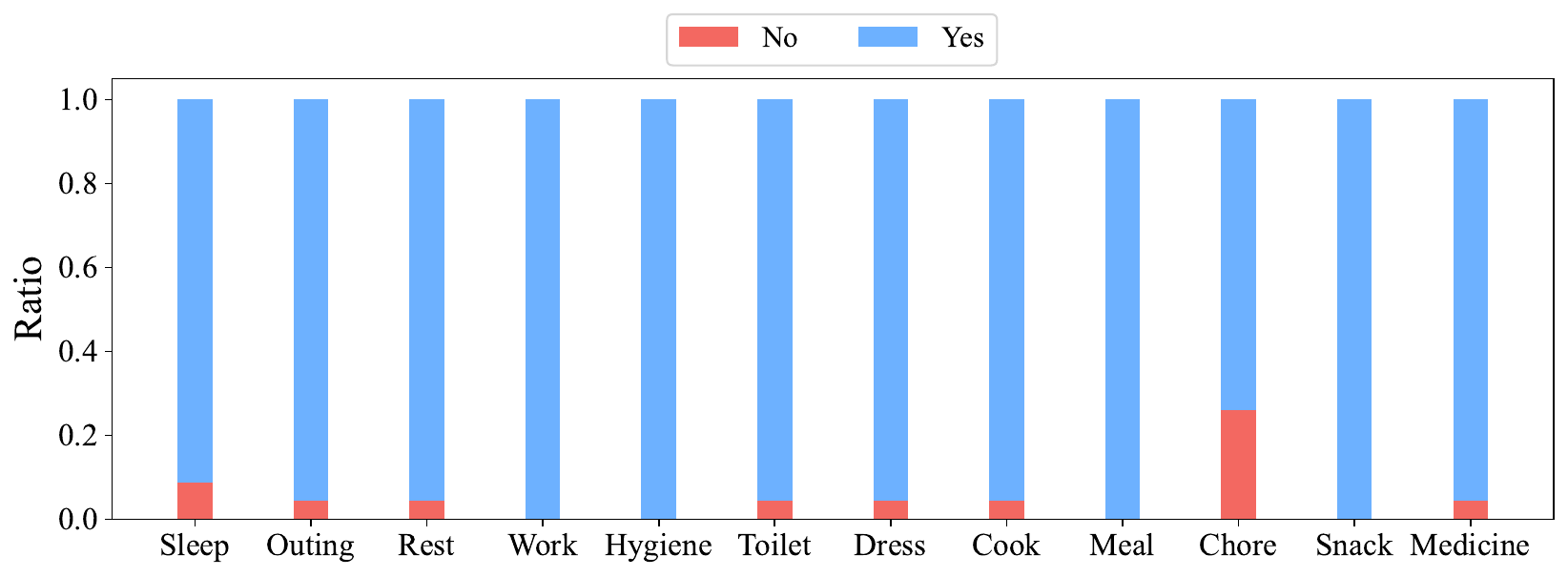}
    \caption{The majority of users responded that the proposed filtering criteria for each activity are suitable for use as the minimum conditions to evaluate the activity's plausibility.}
    \label{fig:Human survey results}
\end{figure*}

\subsection{Preprocessing}
To enhance the dataset’s usability for training, we applied the following preprocessing steps.  
First, the duration of the first and last daily activities, typically sleep, was adjusted to span the entire 24-hour period from 00:00:00 to 23:59:59.  
Next, for segments labeled as \textit{Other activity}, the initial half was replaced with the preceding activity, and the latter half with the subsequent activity. This was also applied to shorter intermittent gaps by extending neighboring activities.  
To mitigate class imbalance, the original 42 and 46 activity categories were consolidated into 12 major activities by grouping semantically similar activities.

The dataset was then reformatted into uniform sequences of 86,400 seconds (24 hours). Due to GPU memory constraints, the sequences were further segmented into 5-minute intervals, with each interval labeled based on the predominant activity. This resulted in sequences of 288 intervals per day, making training computationally feasible.

In the CASAS smart home dataset, 42 activity classes in Apartment dataset and 46 activity classes in Home dataset were used to label the human daily activities. During the preprocessing, we consolidated 42 and 46 activity categories into 12 major activities to enhance the learning by grouping similar activities, as shown in Table~\ref{tab:majorclassification}.
For example, the major activity class ``Sleep" encompasses specific behaviors related to sleeping, such as ``sleep", and ``sleep of bed". 


\begin{figure*}[t!]
    \centering 
    \includegraphics[height=0.491\textwidth]{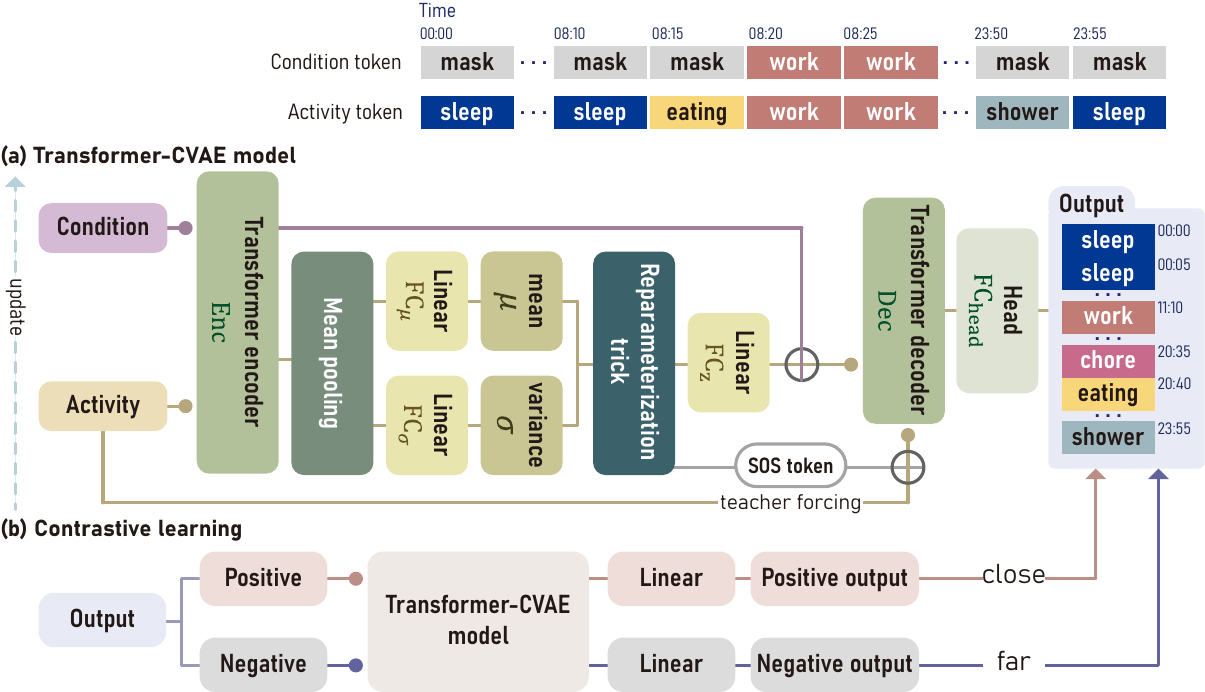}
    \caption{The workflow of ORACLE framework.}
    \label{fig:model_architecture}
\end{figure*}

\subsection{Data Filtering}
\label{sec:Data-Filtering}

Following preprocessing, we applied a filtering step to remove unrealistic activity patterns based on duration and frequency constraints. Filtering criteria were established through a combination of:  
1) Common sense reasoning using ChatGPT~\cite{achiam2023gpt},  
2) A user survey involving 23 participants (16 males, 7 females, aged 22-26, $\mu$ = 23),  
3) Expert validation by the authors.

To ensure high-quality training data, activity sequences that deviated significantly from typical human behavior were excluded. After filtering, the dataset retained approximately \textit{57,216 hours} of recorded activity from the Apartment dataset and \textit{26,112 hours} from the Home dataset. The filtered data was then split into training, validation, and test sets using an 8:1:1 ratio, ensuring a robust evaluation framework.

By emphasizing \textit{total recorded hours} rather than individual sequence counts, we highlight the scale and depth of human activity patterns captured in the dataset. 


Filtering criteria were delineated from the ChatGPT's knowledge, human knowledge, and the author's intuition.  
Exploiting GPT's knowledge base, we identified ``Duration" and ``Frequency" as universal numerical parameters for defining 12 activities and determined their desired values considering the plausible human daily activities. 
The text prompts and GPT outputs are presented in Appendix.

Subsequently, a survey involving human participants was conducted to examine the plausible duration and frequency for each of the 12 activities. Integrating insights from both GPT outcomes and human responses, the authors established the filtering criteria utilizing their intuitive judgment.

To validate these criteria, a subsequent survey assessed human consensus regarding their appropriateness for evaluating the plausibility of activities. As shown in Figure~\ref{fig:Human survey results}, the participants predominantly affirmed the suitability of these criteria for such evaluations.
The survey questions are presented in Appendix. These results suggest that the proposed criteria are reasonable as conservative minimum plausibility checks, and that they capture broadly acceptable constraints rather than idiosyncratic rules tied to a specific individual or household.
Consequently, the criteria provide a more general and reproducible basis for filtering and evaluation across different sequences and settings within our indoor daily-routine domain.

\section{Methodology}

The ORACLE framework, as illustrated in Figure~\ref{fig:model_architecture}, incorporates Transformer architecture, CVAE, and contrastive learning techniques. 
The utilization of Transformer architecture targets the issue of class imbalance within daily activity datasets, where activities like sleeping may dominate others like restroom usage. 
This architecture aims to appropriately weigh activities based on their occurrence rates, enhancing the model's capability to produce balanced and plausible daily plans.

Furthermore, the CVAE is employed within ORACLE to enable the generation of daily activity plans under two scenarios: scenarios with NPCs having predetermined activities and those without any pre-scheduled activities.
This flexibility allows ORACLE to tailor daily plans that seamlessly adapt to various contexts. Our activity plans are represented as discrete tokens over 12 activity classes, and the generation objective focuses on producing coherent token sequences under optional masking constraints.
In this discrete, long-horizon setting, a Transformer-based CVAE provides a natural way to model stochasticity and controllability in sequence space, while remaining simple and stable to train on limited data.
Although diffusion models are a promising alternative, they typically require additional design choices (e.g., discrete diffusion formulations or continuous relaxations) and higher computational cost; thus we adopt a Transformer-CVAE as a more direct fit to our tokenized activity representation.

Lastly, contrastive learning is exploited to enhance the learning efficacy and quality of generated samples.
By distinguishing between similar and dissimilar samples, contrastive learning facilitates a more nuanced understanding and representation of the data, contributing significantly to the model's overall performance. 

\subsection{Transformer-CVAE model}
The inputs for training consist of activity and condition tokens. 
Activity tokens form a sequence of 288-length daily activities \( A = \{a_1, a_2, \ldots, a_{288}\} \), with each token \( a_i \) corresponding to a distinct activity class spanning a 5-minute duration.
These tokens are processed through the Transformer encoder \(Enc\), as defined by:

\begin{equation}
Enc(A) = \text{SelfA}(A) \in \mathbb{R}^{288 \times C},
\label{eq:SelfAttn}
\end{equation}
where SelfA(·) represents the self-attention blocks within the Transformer encoder, and $C=768$ represents the embedding dimension. 
The self-attention mechanism, adept at capturing both local and global dependencies within the sequence, is effective for long-sequence management. 
Then, mean-pooling is applied to the encoder's output to produce a single 768-dimensional vector $\bar{V}(A)$, followed by a linear projection to delineate the latent representations into their \textit{mean} ($\mu$) and \textit{variance} ($\sigma$) via \(FC_{\mu}(\cdot)\) and \(FC_{\sigma}(\cdot)\) layers:

\begin{equation}
    \begin{gathered}
        \bar{V}(A) = MeanPooling(Enc(A)), \\
        FC_{\mu}(\bar{V}(A)) = \{\mu_{j} \mid j \in [1, 2, \ldots, C_0]\}, \\ 
        FC_{\sigma}(\bar{V}(A)) = \left\{\log \sigma_{j}^2 \mid j \in [1, 2, \ldots, C_0]\right\},
        \label{eq:FC}
    \end{gathered}
\end{equation}
where $C_0$ represents the embedding dimension, and \(FC_{\mu}\) and \(FC_{\sigma}\) represent two linear projection layers that transition data from \(\mathbb{R}^{C}\) to \(\mathbb{R}^{C_0}\).
Although $C_0$ is potentially smaller than $C$, the model performance is optimized when $C_0 = C$ in our internal test. 

Subsequently, we employ a reparametrization technique to sample the latent variable \(z_j\), constructing a set of latent variables \(Z = \{z_1, z_2, \ldots, z_{C_0}\}\):
\begin{equation}
  z_{j} = \mu_{j} + \sigma_{j} \cdot \epsilon,
  \label{eq:KLContinuous}
\end{equation}
where \(\epsilon \sim \mathcal{N}(0, 1)\). 
To ensure that the distribution of the latent variables remains close to a standard normal distribution, thereby facilitating efficient and meaningful latent space exploration, Kullback-Leibler(KL) regularization, represented as the regularization loss term $L_{reg}$, is applied with a weighting of $1 \times 10^{-5}$: 
\begin{equation}
L_{reg}(\bar{V}(A)) = \frac{1}{C_0} \sum_{j=1}^{C_0} \frac{1}{2} \left( \mu_{j}^2 + \sigma_{j}^2 - \log(\sigma_{j}^2) - 1 \right).
\label{eq:Lreg}
\end{equation}

Another training input is the sequence of condition tokens \(M = \{m_{1}, m_{2}, \ldots, m_{288}\}\) where each $m_i$ represents whether the corresponding activity token $a_i$ is masked. For example, in scenarios where the activities of $a_3$ and $a_4$ are pre-scheduled, the token $m_3$ and $m_4$ denote an unmasked state, whereas the remaining elements in $M$ indicate a masked state. 

To optimize parameter efficiency within the model, we employ the same Encoder \(Enc\) embedding both activity and condition tokens. 
Then, the latent variable set \(Z\) undergoes transformation via a linear layer \(FC_z(\cdot)\) and is then concatenated with the condition token embeddings \(Enc(M)\):
\begin{equation}
M' = \text{Concat}(FC_z(Z), Enc(M)), \label{eq:conditionconcat}
\end{equation}
$M'$ serves as both the key and value within the Transformer decoder's cross-attention layer:
\begin{equation}
 Dec(A, M) = \text{CrossA}(E_{MSA}(A), M') \in \mathbb{R}^{288 \times C},
\label{eq:ConcatAndCrossAttn}
\end{equation}
where \(E_{MSA}(A)\) represents the output from masked self-attention blocks, facilitating auto-regressive decoding. 
The notation CrossA(·) refers to cross-attention blocks within the Transformer decoder architecture. 

Then, the probabilities for each of the 12 possible activities, denoted as $P_{act}$, are calculated through the last fully connected layer \(FC_{head}\):
\begin{equation}
 P_{act} = FC_{head}(Dec(A, M)).
\label{eq:ConcatAndCrossAttn}
\end{equation}

This model training follows the Maximum Likelihood Estimation (MLE) methodology, leveraging \(FC_z(Z)\) as the start of the sequence (SOS) token. The primary focus is on reconstructive accuracy, which is enhanced by minimizing the negative log-likelihood through reconstruction loss term $L_{recon}$:

\begin{equation}
 L_{recon} = -\sum_{A \in D} \log P(A | A, M; \theta),
\label{eq:NLL}
\end{equation}
where \(\theta\) denotes the model parameter of the model, \(P\)(·) denotes the probabilistic output of the ORACLE model, and \(D\) denotes a training dataset.

\begin{table*}[tb]
  \centering
  \setlength{\tabcolsep}{2pt}
  \footnotesize
  \begin{tabular}{
    C{0.076\linewidth}
    C{0.067\linewidth}
    C{0.050\linewidth}
    C{0.063\linewidth}
    C{0.086\linewidth}
    C{0.086\linewidth}
    C{0.086\linewidth}
    C{0.090\linewidth}
    C{0.047\linewidth}
    C{0.100\linewidth}
    C{0.100\linewidth}
    C{0.100\linewidth}
  }
    \toprule
    \multirow{2}{*}{Datasets} & \multirow{2}{*}{Model} & \multirow{2}{*}{WD $\downarrow$} & LLM & REAL- & Distinct-10- & Distinct-15- & \multirow{2}{*}{SAM $\downarrow$} & SAM- & REAL- & Distinct-10- & Distinct-15- \\
    & & & score $\uparrow$ & Random $\uparrow$ & Random $\uparrow$ & Random $\uparrow$ & & 90 $\downarrow$ & Condition $\uparrow$ & Condition $\uparrow$ & Condition $\uparrow$ \\
    \midrule
    \multirow{4}{*}{\makecell{Apartment \\ dataset}} 
    & Bowman  & 0.6263          & 0.4750          & 0.4715          & 0.1899          & 0.3262          & -               & -               & -               & -               & -               \\
    & Vaswani & -               & -               & -               & -               & -               & 0.5423          & 1.0491          & 0.3430          & 0.1760          & 0.2809          \\
    & Koushik & -               & -               & -               & -               & -               & \textbf{0.3390} & 0.7099          & 0.6131          & 0.2633          & 0.4032          \\
    & ORACLE  & \textbf{0.4835} & \textbf{0.5240} & \textbf{0.8416} & \textbf{0.2106} & \textbf{0.3401} & 0.4218          & \textbf{0.6899} & \textbf{0.9423} & \textbf{0.2831} & \textbf{0.4327} \\
    \midrule
    \multirow{4}{*}{\makecell{Home \\ dataset}} 
    & Bowman  & 1.0470          & 0.4600          & 0.1267          & 0.1699          & 0.2782          & -               & -               & -               & -               & -               \\
    & Vaswani & -               & -               & -               & -               & -               & 0.5976          & 1.0960          & 0.2000          & 0.2305          & 0.3553          \\
    & Koushik & -               & -               & -               & -               & -               & \textbf{0.4437} & 0.9157          & 0.5666          & \textbf{0.3396} & \textbf{0.5109} \\
    & ORACLE  & \textbf{0.8256} & \textbf{0.5137} & \textbf{0.7433} & \textbf{0.1969} & \textbf{0.3414} & 0.4495          & \textbf{0.7647} & \textbf{0.9133} & 0.3276          & 0.4973          \\
    \bottomrule
  \end{tabular}
  \caption{Performance benchmark of ORACLE against three existing models.}
  \label{tab:generation_evaluation}
\end{table*}

\begin{table*}[tb]
  \centering
  \setlength{\tabcolsep}{2pt}
  \footnotesize
  \begin{tabular}{
    C{0.076\linewidth}
    c
    C{0.050\linewidth}
    C{0.063\linewidth}
    C{0.086\linewidth}
    C{0.086\linewidth}
    C{0.086\linewidth}
    C{0.047\linewidth}
    C{0.047\linewidth}
    C{0.100\linewidth}
    C{0.100\linewidth}
    C{0.100\linewidth}
  }
    \toprule
    \multirow{2}{*}{Datasets} & \multirow{2}{*}{Model} & \multirow{2}{*}{WD $\downarrow$} & LLM & REAL- & Distinct-10- & Distinct-15- & \multirow{2}{*}{SAM $\downarrow$} & SAM- & REAL- & Distinct-10- & Distinct-15- \\
    & & & score $\uparrow$ & Random $\uparrow$ & Random $\uparrow$ & Random $\uparrow$ & & 90 $\downarrow$ & Condition $\uparrow$ & Condition $\uparrow$ & Condition $\uparrow$ \\
    \midrule
    \multirow{3}{*}{\makecell{Apartment \\ dataset}}
    & ORACLE -C & 0.7161          & 0.4879          & 0.5942          & 0.1582          & 0.2518          & 0.4233          & 0.6964          & 0.9058          & 0.2662          & 0.4098          \\
    & ORACLE -T & 0.5281          & 0.4849          & 0.4985          & 0.1861          & 0.3211          & 0.4278          & \textbf{0.6867} & 0.7153          & 0.2731          & 0.4237          \\
    & ORACLE    & \textbf{0.4835} & \textbf{0.5240} & \textbf{0.8416} & \textbf{0.2106} & \textbf{0.3401} & \textbf{0.4218} & 0.6899          & \textbf{0.9423} & \textbf{0.2831} & \textbf{0.4327} \\
    \midrule
    \multirow{3}{*}{\makecell{Home \\ dataset}}
    & ORACLE -C & 1.2898          & 0.4220          & 0.4833          & 0.1300          & 0.2188          & 0.4744          & 0.8262          & 0.8500          & 0.2984          & 0.4482          \\
    & ORACLE -T & 1.0463          & 0.4430          & 0.4200          & 0.1719          & 0.2719          & 0.4813          & 0.7896          & 0.5633          & \textbf{0.3300} & 0.4934          \\
    & ORACLE    & \textbf{0.8256} & \textbf{0.5137} & \textbf{0.7433} & \textbf{0.1969} & \textbf{0.3414} & \textbf{0.4495} & \textbf{0.7647} & \textbf{0.9133} & 0.3276          & \textbf{0.4973} \\
    \bottomrule
  \end{tabular}
  \caption{Ablation study results.}
  \label{tab:ablation_study}
\end{table*}

\subsection{Contrastive Learning}
Our approach is anchored in a Transformer-CVAE structure, designed for a reconstruction task that processes input data to generate new data closely mimicking the original. 
While contrastive learning is widely used for representation learning, it has rarely been tailored to long-horizon \emph{discrete daily schedule generation} where realism must be judged at the sequence level (24 hours, 288 tokens) and the model is trained without task-specific pretraining.
In ORACLE, we adapt contrastive learning to this setting by \emph{defining positives and hard negatives in the space of generated schedules} using plausibility criteria (Table~\ref{tab:Criteria}), rather than relying on instance identity or data augmentation as in standard contrastive pipelines.
This design explicitly pushes the decoder away from near-miss but unrealistic routines, providing weak supervision aligned with our planning objective and serving as the key ingredient that makes contrastive learning effective for NPC routine synthesis.

Here, we employ contrastive learning to consider the divergence between generated data \(A_{gen}\), either semantically or structurally, even when they remain within close proximity to the original data's distribution in the latent space.
In cases where the generated data \(A_{gen}\) aligns with the evaluation criteria as presented in Table~\ref{tab:Criteria}, it is classified as a positive sample \(A_{pos}\); otherwise, it is deemed a negative sample \(A_{neg}\).
This scheme facilitates the model's enhanced learning from ``hard" negative samples.
Loss functions for contrastive learning loss are defined as follows:

\begin{equation}
    \begin{gathered}
    L_{positive} = 1 - sim(Dec(A), Dec(A_{pos})), \\
    L_{negative} = sim(Dec(A), Dec(A_{neg}))^2 \\
    L_{contrastive} = L_{positive} + L_{negative}.
    \end{gathered}
\label{eq:contrastive}
\end{equation}

Implementing contrastive learning introduces non-ground truth tokens during training, reducing the gap between training and inference. This strategy alleviates the exposure bias commonly encountered in autoregressive models using teacher forcing.

Furthermore, the generation of positive and negative samples as part of the contrastive learning process not only augments the data diversity but also acts as a mechanism for knowledge injection.
Such an approach is particularly beneficial for models like ours that commence without a pre-trained foundation, lacking the extensive knowledge typically derived from large datasets. 

\section{Experiments}
\subsection{Implementation details}
Our model uses the BERT-base-uncased architecture with 768 hidden units, 12 attention heads, 12 transformer layers, and Gaussian Error Linear Unit (GELU) activation.
This setup balances computational efficiency with the ability to capture complex data patterns, leveraging the proven effectiveness of BERT-base in sequential data processing. All experiments were conducted using a single RTX 3090 GPU.

\subsection{Comparative Evaluation}
\label{sec:quantity evaluation}
This section benchmarks ORACLE against three models in two scenarios: random generation and conditional generation.
For a fair comparison with prior activity-sequence generation approaches on CASAS-style data, we include representative LSTM-based and Transformer-based baselines that are trained/evaluated on the same preprocessed CASAS splits used in our experiments.
Specifically, Bowman et al.~\cite{bowman2016generating} serves as an LSTM-based latent generative baseline for unconditional generation, while Vaswani et al.~\cite{NIPS2017_transformer} (Transformer) and Koushik et al.~\cite{koushik2023activity} (biLSTM) provide strong non-latent sequence-modeling baselines for the masked completion setting.
In the random generation scenario, where condition tokens are fully masked, ORACLE is compared to the VAE-LSTM model by Bowman~\cite{bowman2016generating}. The LSTM’s sequential processing constraints limit its ability to handle long sequences and complex activity relationships, where ORACLE shows advantages.
In the conditional generation scenario, where activities are partially masked, ORACLE is compared to models by Vaswani~\cite{NIPS2017_transformer} and Koushik~\cite{koushik2023activity}. These models lack generative flexibility, producing identical outputs for the same inputs, which limits diversity. ORACLE, by contrast, excels in generating more diverse and contextually appropriate plans, even under data sparsity.
All evaluations were conducted using a test dataset, with the generation process repeated 10 times for statistical robustness.

\subsection{Metrics}
For the random generation scenario, our model's performance is evaluated using the Wasserstein distance (WD), LLM score, REAL-Random, and Distinct-10/15-random. WD measures differences between distributions while the LLM score evaluates the realism and human likeness on a scale from 0 to 100, leveraging the ChatGPT's comprehensive knowledge base.
To ensure consistency with other metrics, the LLM score is normalized by dividing by 100.
The used text prompts and their outcomes are presented in Appendix.
REAL-Random score quantifies the proportion of randomly generated outputs that meet the selection criteria detailed in Table~\ref{tab:Criteria}, indicating how well the model adheres to realistic constraints. 
Distinct-10/15-Random provides our primary quantitative \emph{diversity} analysis by measuring the uniqueness of long-range \textit{n}-grams~\cite{li-etal-2016-diversity} across multiple generations, with \textit{n}=10 and 15.
Higher Distinct-10/15 indicates that the model produces a broader set of non-redundant activity subsequences (i.e., less mode collapse), which is particularly important for evaluating the diversity benefit of latent-variable generation.

For the conditional generation scenario, our model's performance is evaluated using the Sequence Alignment Method (SAM) distance, REAL-Condition, and Distinct-10/15-Condition.
SAM distance assesses how closely the generated sequence matches the ground truth by applying penalties across the entire sequence of 288 elements: ``1" for additions or deletions, and ``2" for substitutions. Here, a higher penalty indicates less similarity.
These cumulative penalties are then normalized by the sequence length.
REAL-Condition and Distinct-10/15-Condition metrics are identical to the REAL-Random and Distinct-10/15-Random metrics used in the random generation scenario, but they are specifically tailored to evaluate model performance under our conditional generation scenarios.

\begin{figure*}[t!]
    \centering 
    \includegraphics[height=0.35\textwidth]{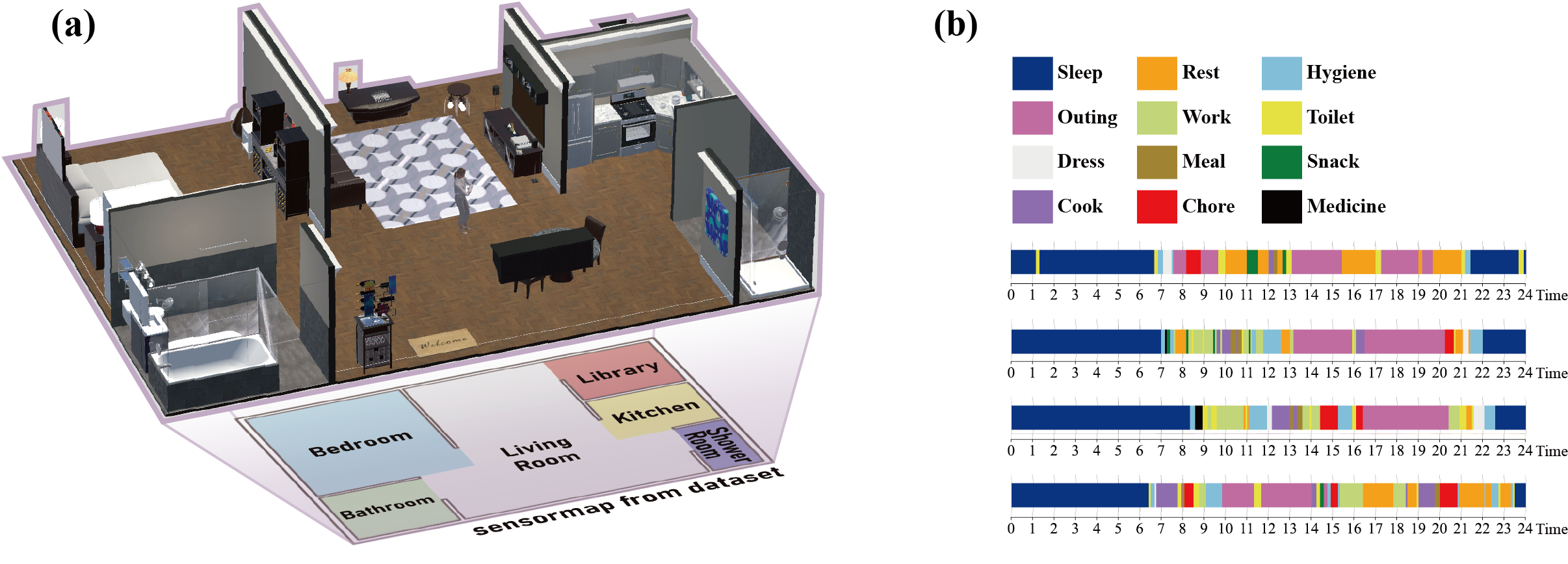}
    \caption{(a) Unity virtual environment where an NPC executes activities.  (b) Samples of the visualized plan.}
    \label{fig:user_study_visualization}
\end{figure*}

\begin{figure}[t!]
    \centering 
    \includegraphics[width=0.50\textwidth]{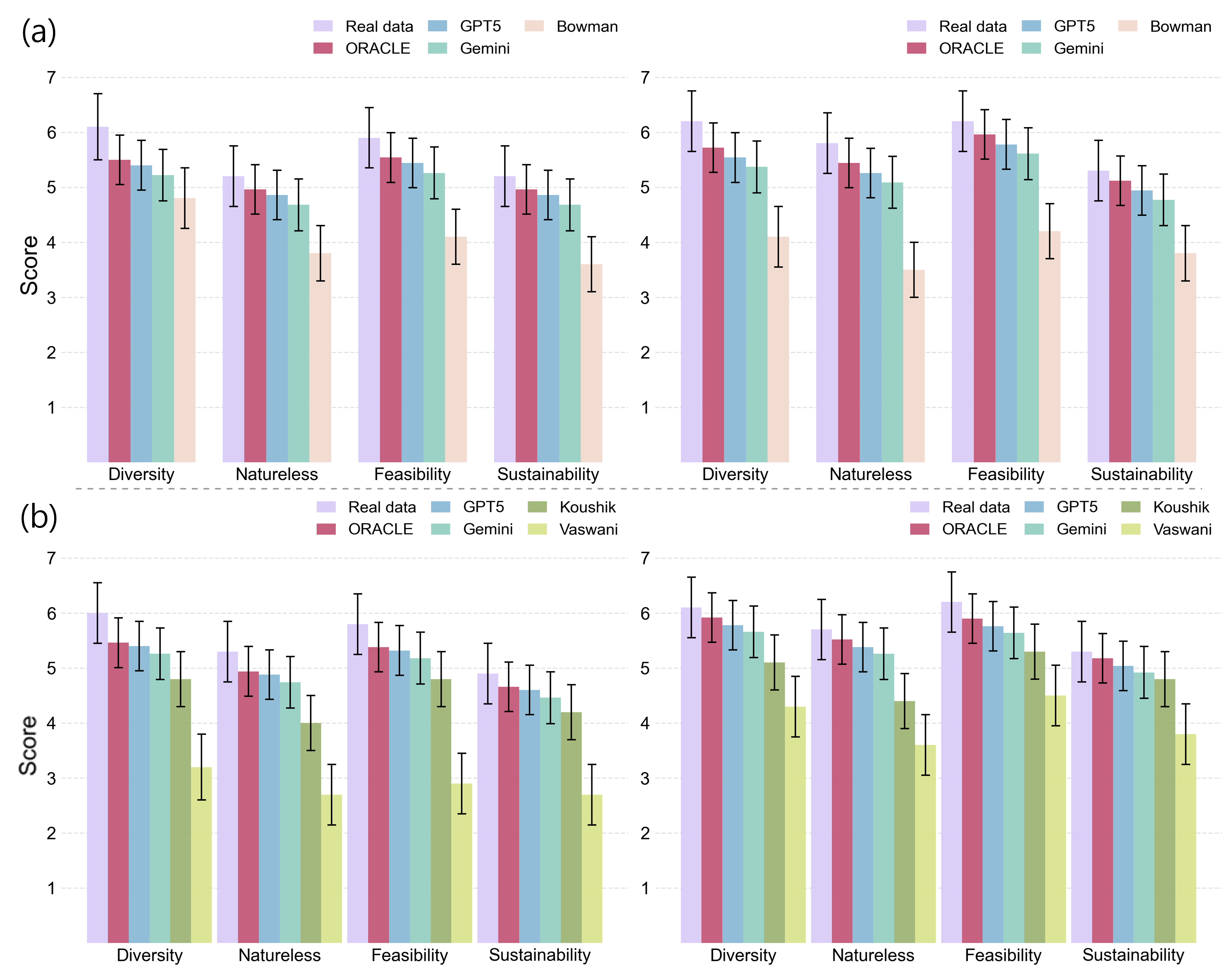}
    \caption{User study evaluation results. (a) Random generation scenario results in Apartment and Home datasets. (b) Conditional generation scenario results in Apartment and Home datasets. Real data is sampled from the test dataset.}
    \label{fig:user_study_stats}
\end{figure}

\begin{figure}[t!]
    \centering 
    \includegraphics[width=0.48\textwidth]{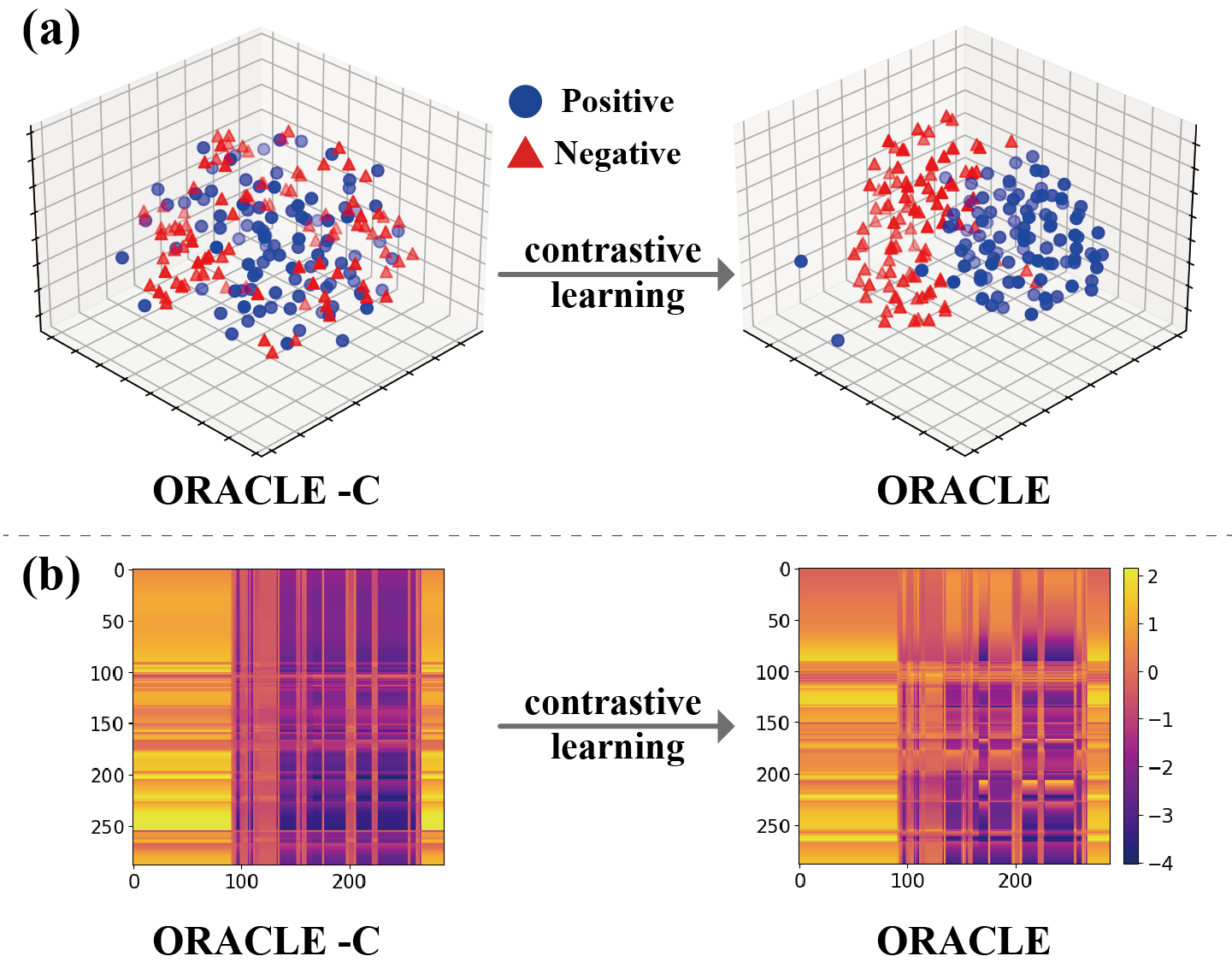}
    \caption{(a) Latent visualization of ORACLE and ORACLE -C. (b) Attention map of ORACLE and ORACLE -C.}

\label{fig:latent_spacespace_attention}
\end{figure}

\begin{figure*}[ht]
\centering
\begin{minipage}{0.49\textwidth}
    \centering
    \includegraphics[width=\textwidth]{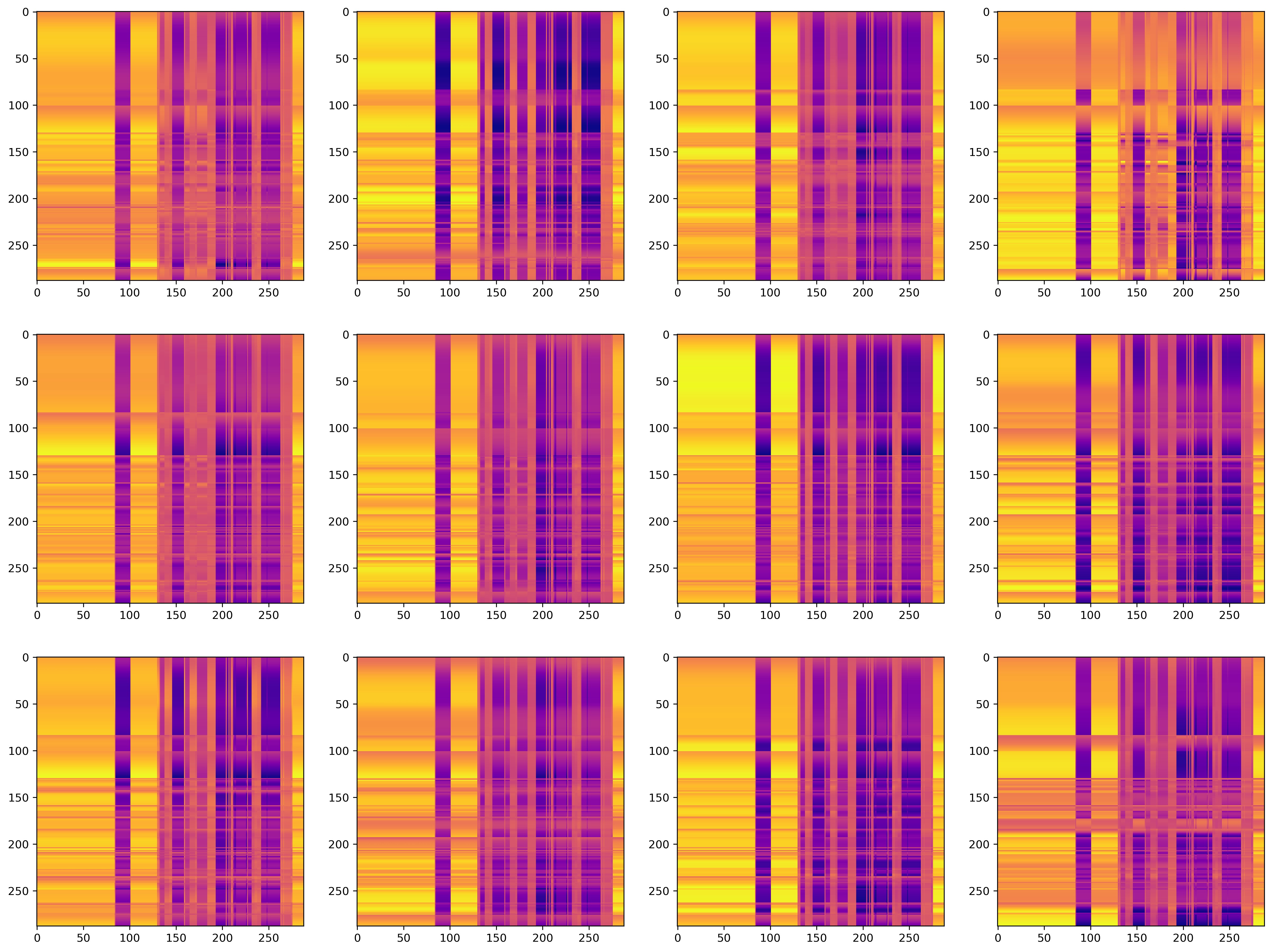}
    \textbf{(a)} Attention maps for ORACLE -C.
\end{minipage}
\hfill
\begin{minipage}{0.49\textwidth}
    \centering
    \includegraphics[width=\textwidth]{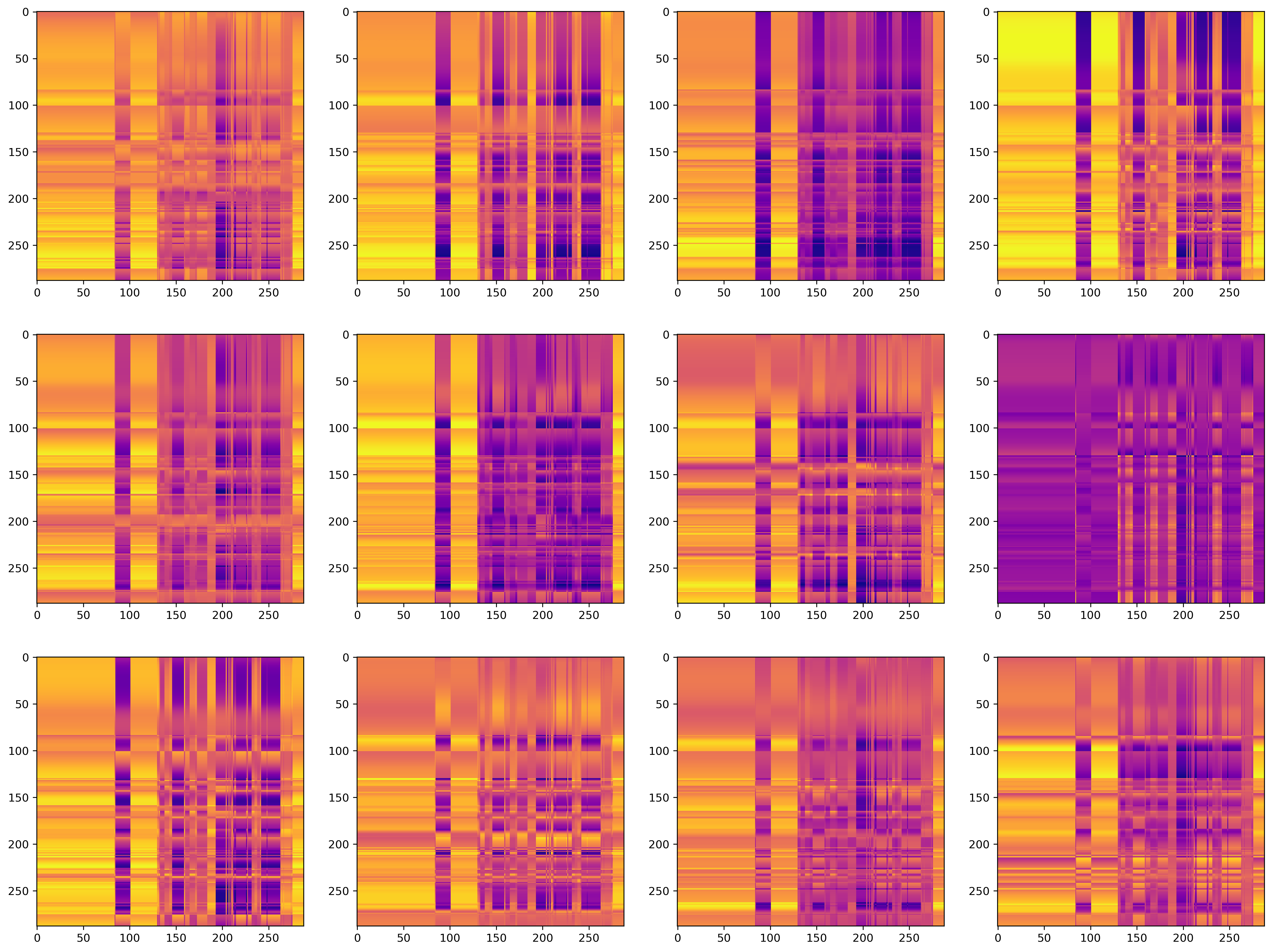}
    \textbf{(b)} Attention maps for ORACLE.
\end{minipage}
\caption{Attention map visualizations of ORACLE and ORACLE -C.}
\label{fig:attention-maps}
\end{figure*}

\begin{figure*}[t!]
    \centering 
    \includegraphics[width=0.80\textwidth]{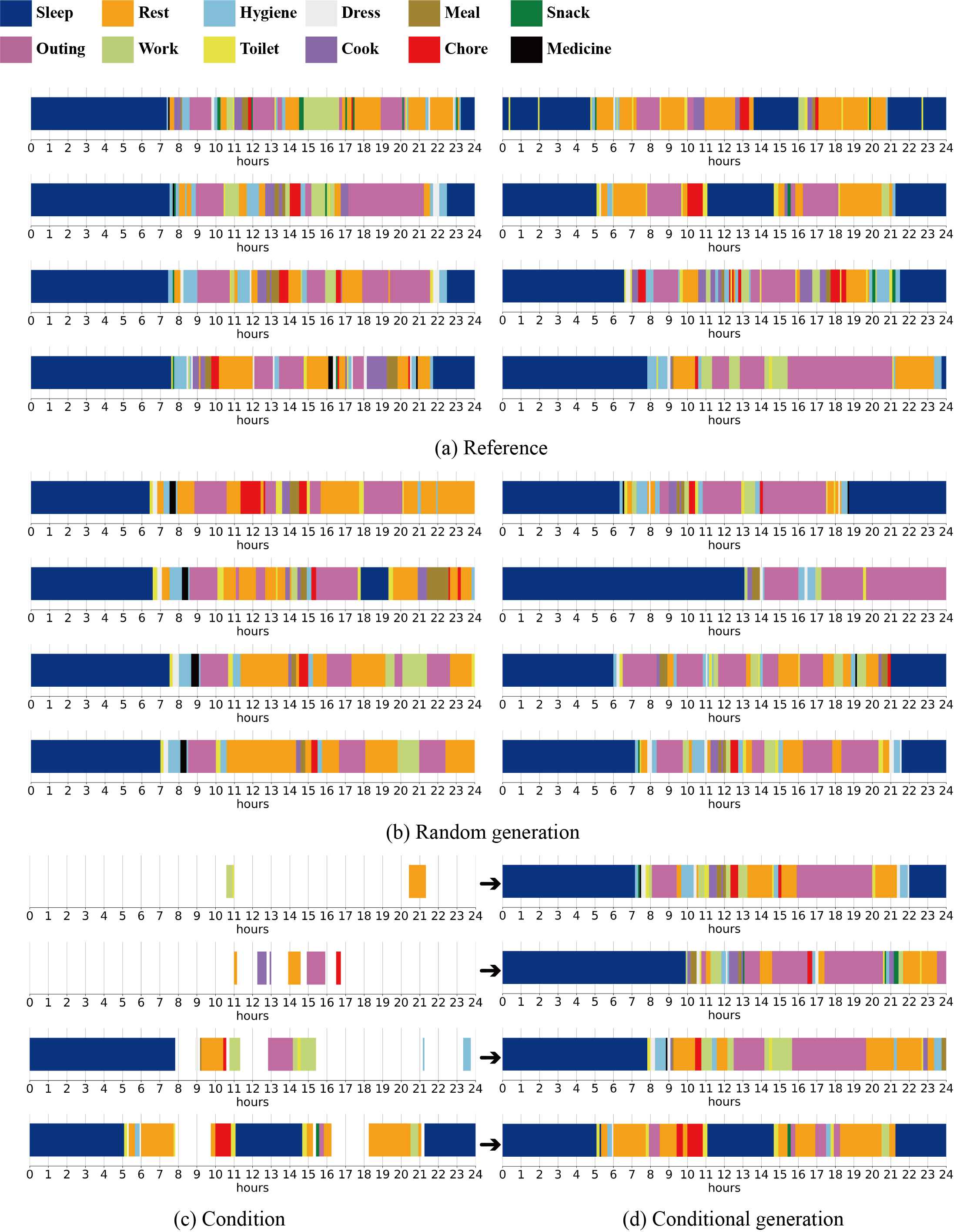}
    \caption{The visualization of daily activities generated by our model.}
    \label{fig:visualization_8}
\end{figure*}

\subsection{Evaluation Results}

The results of the evaluation are presented in Table~\ref{tab:generation_evaluation}. For the random generation scenario, ORACLE outperforms the hybrid LSTM-VAE model by Bowman.   
Firstly, a lower WD signifies a reduced discrepancy between two probability distributions. In generative model contexts, this suggests that the model's generated data distribution closely mirrors the true data distribution, implying more accurate learning of the data's distribution.
Secondly, a higher LLM score signifies that the generated outputs are more plausible.
This indicates that the model effectively captures the relationships between activity classes, leading to daily activity plans that appear more human-like. Additionally, ORACLE achieves higher REAL-Random, Distinct-10-Random, and Distinct-15-Random scores, demonstrating that the generated outputs are not only realistic but also diverse. The high REAL-Random score indicates that ORACLE generates sequences that better meet the criteria in Table~\ref{tab:Criteria}, while the higher Distinct-10-Random and Distinct-15-Random scores reflect the model’s ability to produce a wide variety of unique and non-repetitive sequences.

For the conditional generation scenario, the comparative performance of ORACLE is mixed. 
In terms of SAM distance, ORACLE outperforms the Transformer model by Vaswani yet falls short compared to the bidirectional LSTM-based model by Koushik. 
This is attributed to the ORACLE's generative nature, which promotes diversity, consequently placing ORACLE at a comparative disadvantage when evaluated using SAM distance.
Nevertheless, when the masking ratio exceeds 90\%, as assessed by SAM-90, ORACLE is observed to outperform the aforementioned two models.
Moreover, ORACLE generally exhibits higher REAL-Condition, Distinct-10-Condition and Distinct-15-Condition scores, indicating that it generates sequences that are not only more aligned with the given conditions but also maintain greater diversity, producing a broader range of unique sequences under specific constraints.

Additionally, we examined whether ORACLE simply memorizes and reproduces training samples by performing a kNN analysis between generated sequences and the training set. Since each day is discretized into a 288-length sequence with 5-minute activity tokens, we use the Hamming mismatch (the number of different tokens) as the distance. Across all settings, we observed zero exact matches (distance = 0), indicating that ORACLE does not directly copy any training sequence. Concretely, for the Apartment dataset, ORACLE random generation yields a top-1 nearest-neighbor mismatch of 99.50/288 (34.55\%) on average (median 98, minimum 67), and a top-5 average mismatch of 105.38/288 (36.59\%). For Apartment conditional generation, the top-1 mismatch is 96.47/288 (33.50\%) on average (median 98, minimum 40), with a top-5 average mismatch of 102.14/288 (35.46\%). For the Home dataset, ORACLE random generation shows a top-1 mismatch of 117.50/288 (40.80\%) on average (median 118, minimum 94) and a top-5 average mismatch of 123.37/288 (42.84\%). Home conditional generation produces comparatively closer sequences, with a top-1 mismatch of 62.30/288 (21.63\%) on average (median 54, minimum 7), which is expected because conditional generation can converge to training-distribution days that satisfy the given constraints; nevertheless, no exact reproductions occur (distance = 0 for all samples).

\subsection{User Evaluation}

This section evaluates the subjective quality of daily activity plans generated by ORACLE in comparison to three benchmark models, as discussed in section \ref{sec:quantity evaluation}. The evaluation adopts the same dual scenarios of random generation and conditional generation. 

Eleven participants (7 males, 4 females, age range 22-29, $\mu=24.36$, $\sigma=2.062$) participated in the study.
Participants evaluated a total of 36 data samples, randomly selected and visualized, combining both the random generation and conditional generation scenarios in both Apartment and Home datasets.
The evaluation involved data visualization and a 2.5-minute virtual environment simulation for each sample.
The virtual environment is illustrated in Figure~\ref{fig:user_study_visualization}(a), and the data visualization is shown in Figure~\ref{fig:user_study_visualization}(b).
They then rated four research questions on a 7-point Likert scale: ``\textit{Diversity}: Are these activities diverse and varied?", ``\textit{Naturalness}: Do these activities resemble how a person would do within a day?", ``\textit{Feasibility}: Do you think these activities could be done by someone?", and ``\textit{Sustainability}: Could you see yourself living comfortably for a week or more following these plans?"

The analysis results of the random and conditional generation scenarios in Apartment and Home datasets are presented in Figure~\ref{fig:user_study_stats}.
When comparing ORACLE to Bowman., ORACLE achieves slightly higher scores across all metrics. 
These findings align with the LLM score's validity tested in section \ref{sec:quantity evaluation} and demonstrate that ORACLE generates more human-like and credible activity plans.
Against the models by Vaswani and Koushik, ORACLE demonstrates superior performance, indicating its proficiency in crafting plausible schedules, even when they are partially pre-scheduled. 
We additionally compared ORACLE with two recent LLMs~\cite{achiam2023gpt, comanici2025gemini} using the same prompt: \textit{“Generate a 24-hour human-like daily behavior sequence in 5-minute increments (288 values total) using {Sleep, Outing, Rest, Work, Hygiene, Toilet, Dress, Cook, Meal, Chore, Snack, Medicine}, output in the format \texttt{HH:MM\textasciitilde HH:MM [activity]}.”} As these LLMs were not trained on activity datasets, only qualitative evaluation was performed, where participants rated ORACLE higher.

Our user study provides a subjective perspective on the generated plans, although its scale and scope are inherently limited. We therefore present it as a complementary evaluation alongside our main quantitative results. In particular, we report plausibility-based REAL scores, distributional similarity via WD, and long-range diversity via Distinct-$n$, which capture different aspects of realism and variability in daily routines. The overall trends observed in the user study are broadly consistent with these objective metrics, providing additional support that ORACLE generates realistic and diverse activity plans.

\subsection{Ablation Study}
We conducted an ablation study to evaluate the impact of contrastive learning and Transformer architecture on ORACLE. 
We compared the standard ORACLE model to a variant without contrastive learning (\mbox{ORACLE -C}) and another with an LSTM instead of a Transformer (\mbox{ORACLE -T}).

Table~\ref{tab:Ablation study} shows that standard ORACLE outperforms all variants, presenting the effectiveness of contrastive learning and Transformer architecture.
To further investigate the effect of contrastive learning, we visualized latent space and attention maps for both ORACLE and ORACLE -C, as depicted in Figure~\ref{fig:latent_spacespace_attention}(a) and (b).

In latent space visualization, test dataset samples were identified as positive samples, whereas generated data failing to meet the established criteria were identified as negative samples.
These samples were then rendered using t-SNE. The visualization revealed overlaps between positive and negative samples in ORACLE -C, indicating potential exposure bias. In contrast, ORACLE's contrastive learning improved sample discrimination, reducing bias and enhancing generalization across diverse data scenarios.

In attention map visualization, test dataset samples were selected at random during the task of reconstruction, focusing particularly on the outputs from the final cross-attention layer's 12 heads.
The visualization revealed ORACLE's maps to be more varied than ORACLE -C, indicating superior generalization to unseen scenarios during training.

\subsection{Further Visualization Results}
Figure~\ref{fig:attention-maps} showcases the attention maps obtained from the 12 heads of the final cross-attention layer. 
Prior to the application of contrastive learning, attention maps exhibit a uniform focus. Conversely, after the incorporation of contrastive learning, the model exhibits a more diverse focus, with each head concentrating on different aspects of the data. This indicates that the model analyzes the data from various perspectives at the level of individual heads, suggesting the potential of contrastive learning to enhance the richness and diversity of the outcomes. 

Figure~\ref{fig:visualization_8} provides additional daily activities generated by our model. Analysis of these results reveals that both random and conditional generations exhibit similarities to the reference data, indicating that our model has effectively learned the data's distribution. This resemblance serves as evidence of the model's successful training.

We provide an example video in the supplementary material to demonstrate a minimal proof-of-concept pipeline that maps each predicted activity token to a simple in-engine action and location transition in a Unity apartment scene.
The purpose of this visualization is to show the usability of ORACLE as an activity-plan generator that can drive an NPC controller, rather than to present a complete animation set or high-fidelity character behaviors.
Thus, the limited action variety in the current video reflects the simplicity of the demonstration mapping, not a limitation of the generated plans.
Importantly, the activity-to-action mapping is modular and can be readily extended with richer interaction primitives (e.g., object-level affordances, state machines, motion libraries, or behavior trees), enabling more diverse and convincing executions without changing the underlying planning model.

\section{Conclusion}
In this paper, we introduce ORACLE, a novel approach for generating plausible daily activity plans for NPC utilizing a Transformer-CVAE model with contrastive learning. 
Our experiments confirm that ORACLE outperforms existing models in generating both complete schedules from scratch and enhancing existing ones with additional activities.

ORACLE offers significant potential for applications in enhancing smart home automation, supporting systems for the elderly, aiding in urban planning, and streamlining facility management through its predictive capabilities for personal daily routines. 
Furthermore, the training framework of ORACLE has broader applicability in other domains requiring sequential data generation, such as aligning events in games, simulating urban designs, and automating production processes in manufacturing settings. 

Despite its achievements, our approach has a limitation.
ORACLE primarily facilitates the planning of generalized behaviors inferred from the smart home dataset, thus restricting its capacity to devise personalized activities or predict behaviors occurring outside the home environment. 
The most direct strategy to address this challenge would involve sourcing datasets that encapsulate a wider array of human behaviors. However, the acquisition of such comprehensive data presents significant challenges, rendering this solution impractical. 
An alternative method to overcome this may entail the integration of reinforcement and deep learning strategies to simulate human decision-making processes, thereby accommodating a wider variety of situational decisions.
We will direct our future research to simulate a broader range of human behaviors across various environments.


\section*{Acknowledgement}
This work was supported by the National Research Foundation of Korea (NRF) funded by the
Korean government (MSIT) (project management number granted from IRIS) (No. RS-2025-24802983, 34\%).
This work was also supported by the National Research Foundation of Korea (NRF) grant funded by the Korean government (MSIT) (No. RS-2025-00518643, 33\%).
It was further partly supported by the Institute of Information \& Communications Technology Planning \& Evaluation (IITP)-ICT Creative Consilience Program grant funded by the Korea government(MSIT) (No. IITP-2025-RS-2020-II201819, 33\%).


{\small
\bibliographystyle{cvm}
\bibliography{cvmbib}
}

\clearpage

\begin{table*}[hbtp]
  \centering
    \setlength{\abovecaptionskip}{1pt}
    \setlength{\belowcaptionskip}{-80pt}
    \begin{tabular}{p{\linewidth}}
    \toprule
    Input prompts\\
    \midrule
    I'd like to design daily human activities with a defined set \{Sleep, Outing, Rest, Work, Hygiene, Toilet, Dress, Cook, Meal, Chore, Snack, Medicine\}. I need you to establish numerical parameters for these activities. These parameters must be commonly used for all activities. 
    These parameters should relate to the sequence of activities throughout the day without being tied to specific times. Please recommend parameters and their desired values, drawing on general knowledge about realistic human daily activities.   \\
    \bottomrule
    \\
    \toprule
    GPT outputs \\
    \midrule
    Given the scope of your request, I will focus on two parameters that can be universally applied to the activities listed: Duration and Frequency. These parameters are crucial for planning and can accommodate the variability of daily routines. Below are the recommended values based on general knowledge and assumptions about an average adult's daily activities. Please note that these are guidelines and actual values may vary based on individual needs and lifestyle.\\\\
    
    Sleep; 
    Duration: 7-9 hours (Per night),
    Frequency: 1 time per day
    
    Outing; 
    Duration: 1-4 hours (Per outing),
    Frequency: 1-2 times per day

...\\
    
    Snack;
    Duration: 5-15 minutes (Per snack),
    Frequency: 2-3 times per day

    Medicine;
    Duration: 1-5 minutes (Per session),
    Frequency: 1-3 times per day (based on prescription)\\\\
    
    These values provide a foundational framework for designing a daily schedule, considering a balance between obligations and personal care. Adjustments may be necessary to tailor the schedule to individual needs, preferences, and specific life circumstances.\\
    \bottomrule  
    \hline
  \end{tabular}
  \caption{Input text prompts and GPT outputs for data filtering.}
  \label{tab:LLManalysis}
\end{table*}

\begin{table*}[tb]
  \centering
  \begin{tabular}{p{\linewidth}}
    \toprule
    Question 1: Do you consider the criteria presented to be the minimum standards for sleep time and frequency for a normal person's sustainable life? \\
    Answer 1: People should sleep for at least 5 hours and up to 12 hours per day. \\
    \midrule
    Question 2: Do you consider the criteria presented to be the minimum standards for outing time and frequency for a normal person's sustainable life? \\
    Answer 2: People should go out for up to 12 hours a day. \\
    \midrule
    Question 3: Do you consider the criteria presented to be the minimum standards for rest time and frequency for a normal person's sustainable life? \\
    Answer 3: People take a rest for up to 12 hours a day. \\ 
    \midrule
    Question 4: Do you consider the criteria presented to be the minimum standards for work time and frequency for a normal person's sustainable life? \\ 
    Answer 4: People work for up to 12 hours a day. \\
    \midrule
    Question 5: Do you consider the criteria presented to be the minimum standards for hygiene activities time and frequency for a normal person's sustainable life? \\
    Answer 5: People do hygiene activities such as taking a shower at least once a day, for up to 1 hour and 30 minutes at a time. \\
    \midrule
    Question 6: Do you consider the criteria presented to be the minimum standards for toilet time and frequency for a normal person's sustainable life? \\
    Answer 6: People use the toilet at least once a day, for up to 30 minutes at a time. \\
    \midrule
    Question 7: Do you consider the criteria presented to be the minimum standards for dress time and frequency for a normal person's sustainable life? \\
    Answer 7: People change clothes at least once a day, for up to 1 hour at a time. \\
    \midrule
    Question 8: Do you consider the criteria presented to be the minimum standards for cooking time and frequency for a normal person's sustainable life? \\ 
    Answer 8: People cook for up to 2 hours at a time. \\
    \midrule
    Question 9: Do you consider the criteria presented to be the minimum standards for meal time and frequency for a normal person's sustainable life? \\
    Answer 9: People have meals at least once a day, for up to 2 hours at a time. \\
    \midrule
    Question 10: Do you consider the criteria presented to be the minimum standards for chore time and frequency for a normal person's sustainable life? \\
    Answer 10: People perform household chores such as cleaning and dishwashing at least once a day, for up to 2 hours at a time. \\
    \midrule
    Question 11: Do you consider the criteria presented to be the minimum standards for snack time and frequency for a normal person's sustainable life? \\
    Answer 11: People snack for up to 2 hours at a time. \\
    \midrule
    Question 12: Do you consider the criteria presented to be the minimum standards for medicine time and frequency for a normal person's sustainable life? \\
    Answer 12: People take medication for up to 2 hours at a time. \\
  \bottomrule
  \end{tabular}
  \caption{Survey questions assessing the filtering criteria for each activity.}
  \label{tab:humansurvey}
\end{table*}

\begin{table*}[tb]
  \centering
  
    \begin{tabular}{p{\linewidth}}
    \toprule
    Input prompts\\
    \midrule
    We want to evaluate if the following [number] indoor daily schedules resemble a human's daily routine. Please rate it on a scale from 0 to 100 for each. Tell me the score with further explanation. And find the average of those scores.\\ \\

    00:00\~{}00:05 Rest 00:05\~{}00:15 Chore 00:15\~{}00:20 Dress 00:20\~{}00:35 Hygiene 00:35~08:25 Sleep 08:25\~{}08:30 Toilet 08:30\~{}12:10 Sleep 12:10\~{}12:25 Hygiene 12:25\~{}12:35 Dress 12:35\~{}12:40 Rest 12:40~13:10 Outing 13:10\~{}13:50 Rest 13:50\~{}14:40 Outing 14:40\~{}16:00 Rest 16:00\~{}16:20 Work 16:20\~{}17:55 Rest 17:55\~{}18:10 Hygiene 18:10\~{}18:20 Work 18:20\~{}18:45 Outing 18:45\~{}20:40 Rest 20:40\~{}24:00 Outing \\
    $\cdots$\\
    \bottomrule
    \\
    \toprule
    GPT outputs \\
    \midrule

To evaluate these schedules on a scale from 0 to 100 for how closely they resemble a human’s daily routine, I considered several factors:

Sleep Patterns: Regularity and total duration of sleep.
Activity Distribution: Balanced and realistic distribution of work, chores, meals, and rest.
Consistency: Predictability and normalcy in daily routines, avoiding erratic or excessive repetition of certain activities.
Social and Leisure Activities: Time allocated for social interactions, leisure, and outings.
Overall Realism: How the entire schedule feels when compared to an average person's day. \\ \\

Example Scores: \\
00:00\~{}00:05 Rest 00:05\~{}00:15 Chore 00:15\~{}00:20 Dress 00:20\~{}00:35 Hygiene 00:35\~{}08:25 Sleep...

Score: 40/100: Unusual sleep pattern with significant sleep during morning hours, minimal time for meals and work, and unrealistic hygiene frequency.

00:00\~{}07:10 Sleep 07:10\~{}07:20 Hygiene 07:20\~{}07:25 Snack...

Score: 65/100: A more realistic distribution of activities with decent sleep and work patterns, though still some odd timing for hygiene and meals.

00:00\~{}06:00 Sleep 06:00\~{}06:10 Hygiene 06:10\~{}06:20 Dress...

Score: 70/100: A fairly standard schedule with adequate sleep, hygiene, and work balance, though some activities like multiple short rests are slightly odd.

00:00\~{}09:45 Sleep 09:45\~{}09:55 Hygiene 09:55\~{}10:10 Cook...

Score: 75/100: A good distribution with realistic blocks for sleep, work, and meals, though a bit too much time in outings.

$\cdots$

an estimated average score across all schedules could be around \textbf{50/100}, reflecting a mix of schedules that somewhat resemble human routines 

$\cdots$ \\
  \bottomrule
  \end{tabular}
  \vspace{5pt}
  \caption{Input text prompts and GPT outputs for LLM score.}
  \label{tab:LLMscore}
\end{table*}

\end{document}